\documentclass[conference]{IEEEtran}
\IEEEoverridecommandlockouts
\usepackage{cite}
\usepackage{caption}
\usepackage{amsmath,amssymb,amsfonts}
\usepackage{bm}
\usepackage{diagbox}
\usepackage{algorithm}
\usepackage{algpseudocode}
\usepackage{booktabs}
\usepackage{graphicx}
\usepackage{subfig}
\usepackage{epstopdf}
\usepackage{float}
\usepackage{textcomp}
\usepackage{xcolor}
\usepackage{booktabs}
\usepackage{diagbox}
\usepackage{multirow}

\usepackage{subfig}
\usepackage{extarrows}

\usepackage[dvips]{epsfig}
\usepackage{url}
\usepackage{xspace}
\def\ie{\textit{i.e.}\xspace}
\def\etal{\textit{et al.}\xspace}

\def\eg{\textit{e.g.}\xspace}

\def\BibTeX{{\rm B\kern-.05em{\sc i\kern-.025em b}\kern-.08em
    T\kern-.1667em\lower.7ex\hbox{E}\kern-.125emX}}
\usepackage{color}
\begin{document}
\setlength{\textfloatsep}{3pt}
\title{EdgeLoc: An Edge-IoT Framework for Robust Indoor Localization Using Capsule Networks}

\author{Qianwen~Ye,~
	Xiaochen~Fan{*},
	Gengfa~Fang{*},
	Hongxia~Bie{*},~
	Chaocan~Xiang,
	Xudong~Song
	and~Xiangjian~He
	\thanks{Qianwen Ye and Hongxia Bie are with the School of Information and Communication Engineering, Beijing University of Posts and Telecommunications, Beijing, China (e-mail: biehx@bupt.edu.cn).}
	\thanks{Qianwen Ye, Xiaochen Fan, Gengfa Fang, Xudong Song and Xiangjian He are with the Faculty of Engineering and Information Technology, University of Technology, Sydney, Ultimo NSW 2007, Australia (e-mail: \{qianwen.ye, xiaochen.fan\}@student.uts.edu.au, \{gengfa.fang, xudong.song, xiangjian.he\}@uts.edu.au).}
	\thanks{Chaocan Xiang is with the College of Computer Science, Chongqing University, Chongqing, China (e-mail: chaocan.xiang@gmail.com)}
}

\maketitle

\renewcommand{\thefootnote}{\fnsymbol{footnote}}
\footnotetext[1]{Corresponding authors.}

\begin{abstract}
With the unprecedented demand for location-based services in indoor scenarios,
wireless indoor localization has become essential for mobile users.
While GPS is not available at indoor spaces,
WiFi RSS fingerprinting has become popular with its ubiquitous accessibility.
However, it is challenging to achieve robust and efficient indoor localization with two major challenges.
First, the localization accuracy can be degraded by the random signal fluctuations, which would influence conventional localization algorithms that simply learn handcrafted features from raw fingerprint data.
Second, mobile users are sensitive to the localization delay, but conventional indoor localization algorithms are computation-intensive and time-consuming.
In this paper, we propose EdgeLoc, an edge-IoT framework for efficient and robust indoor localization using capsule networks.
We develop a deep learning model with the CapsNet to efficiently extract hierarchical information from WiFi fingerprint data, thereby significantly improving the localization accuracy.
Moreover, we implement an edge-computing prototype system to achieve a nearly real-time localization process, by enabling mobile users with the deep-learning model that has been well-trained by the edge server.
We conduct a real-world field experimental study with over 33,600 data points and an extensive synthetic experiment with the open dataset, and the experimental results validate the effectiveness of EdgeLoc. The best trade-off of the EdgeLoc system achieves 98.5\% localization accuracy within an average positioning time of only 2.31 ms in the field experiment.
\end{abstract}


\section{Introduction}
With the ubiquitous coverage of wireless networks and the pervasive usage of smart devices, indoor location-based services (ILBSs), such as mobile advertising~\cite{ye2020capsloc}, navigation~\cite{ayyalasomayajula2020deep}, interactive routing~\cite{belmonte2017adaptive}, have become prevailing IoT applications in smart cities.
As a prerequisite of ILBSs, indoor localization, particularly through wireless communications, has become a necessity.
While GPS signal is too sensitive to occlusion (\eg, buildings) and it cannot deliver satisfactory localization results, WiFi~\cite{xiao2016survey} have been intensively utilized for indoor localization,
as it has the widest indoor availability for mobile devices.

\begin{figure}[t]
		\centering
	   \includegraphics[width=0.4\textwidth]{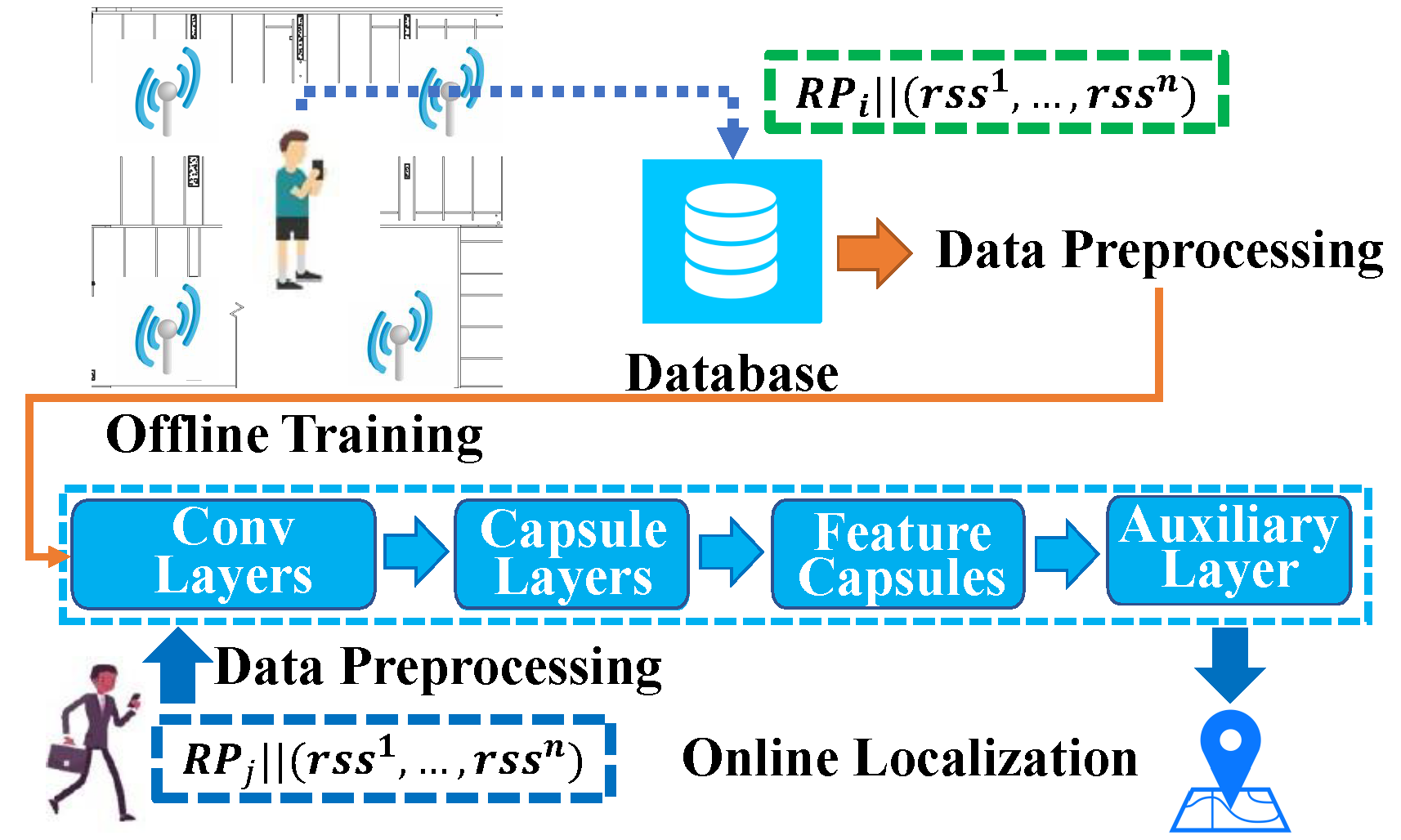}
	   \caption{The scenario of real-time indoor localization with RSS fingerprinting.}
	   \label{scenario}
\end{figure}

In recent years, the RSS-based fingerprinting of WiFi has received numerous efforts for achieving practical and accurate Non-Line-of-Sight (NLoS) localization.
It is assumed that an indoor location can be identified by a unique signal vector of RSS values measured from surrounding WiFi access points (APs).
As illustrated in Fig.~\ref{scenario}, fingerprinting-based localization generally consists of two phases~\cite{yassin2016recent}:
(1) an offline phase for data collection and model construction;
and (2) an online phase for real-time localization by RSS-location mapping.
For the offline phase, the fingerprint dataset is collected at distributed reference points (RPs) across the indoor space. While for the online phase, the localization system would employ different localization algorithms/models to find the best match between the current fingerprint and the corresponding indoor location.

Nevertheless, driven by the huge demand of ILBSs from IoT devices, it is still challenging to realize robust and efficient indoor localization with two major challenges:
\begin{itemize}
  \item First, it is challenging to achieve high-accuracy indoor localization with the random signal fluctuations, which can heavily influence conventional algorithms that simply learn handcrafted features from raw data. In contrast, accurate indoor localization requires effective extractions of reliable features from the fingerprint dataset, as well as effective mapping functions to perform positioning tasks.
  \item Second, it is hard to boost localization speed. Despite that mobile users are particularly latency-sensitive for ILBSs, conventional localization algorithms/models are computation-intensive and time-consuming. Moreover, mobile devices only have limited resources and cannot afford the training process~\cite{RN165}. While remote cloud servers have sufficient resources, there will be transmission delays in long-range communications with them.
\end{itemize}

In this paper, we aim to solve both of the above challenges.
To improve the accuracy, robustness and time efficiency of indoor localization,
we propose to combine deep learning with edge computing.
The key idea is two-fold:
(1) leveraging the deep neural networks to extract hierarchical features from joint representation learning, and thus to improve localization accuracy;
and (2) exploiting low-latency and high-bandwidth access of edge computing, and thereby enabling mobile users with the ability of real-time computation for indoor positioning.
We explicitly explain our idea from the above two perspectives as follows.

First, conventional machine learning algorithms for indoor localization simply learn handcrafted features from RSS fingerprints, and their performance can be easily influenced by the variations and fluctuations of RSS signals.
To efficiently extract high-dimensional representation from complex fingerprint data,
neural network-based architectures have been proposed with the rise of deep learning.
Among these architectures,
CNN shows remarkable performance in processing data in the form of arrays by extracting high-level features with consecutive convolution operations and pooling operations.
However, some valuable spatial information of between-layer neurons could get lost with
pooling operations (\textit{e.g.}, max-pooling) in CNNs.
For indoor localization, losing such information can directly degrade the localization accuracy,
as RSS-based fingerprints are spatially distributed and have strong correlations.
To address this problem, we propose to leverage Capsule Network~\cite{RN104} and propose CapsNet as an alternative to CNNs to efficiently capture the hierarchical structure of the entities in the RSS fingerprinting data.
The Capsules are composed of neurons that use vectors to learn and store feature information,
with each neuron's output representing a different property of the same feature.

Second, in contrast to the centralized cloud computing, edge computing has recently emerged as a new computing paradigm, pushing the computing and storage resources to the logical edge of the network. Therefore, edge computing has become popular for computation-demanding and latency-sensitive mobile applications.
With edge computing, mobile users requesting indoor localization services can benefit from the deployed edge server, where the local database stores all reference data samples and the localization model is already well-trained.
In this way, as long as mobile users get the current RSS fingerprint,
he can directly acquire the optimized model parameters from the edge server,
thus to have a trained localization model computing his location in realtime, at zeros cost, and without communication delay.

The main contributions of this work are summarized as follows.
\begin{itemize}
  \item We propose EdgeLoc, an edge-IoT framework for robust indoor localization using capsule networks. To the best of our knowledge, we are the first to combine edge computing with capsule networks in deep learning for indoor localization. We present the architecture design and systematic dataflow of EdgeLoc and further implement it in a real-world experimental field for indoor localization.
  \item We build CapsNet, the core localization model of EdgeLoc, by leveraging the capsule networks. CapsNet can efficiently capture the hierarchical representations from RSS fingerprinting data, thereby improving the robustness and accuracy of indoor localization with fingerprinting data of RSS.
  \item We conduct extensive studies for the EdgeLoc system in a field experiment and with a large-scale public fingerprinting dataset, respectively. The results demonstrate the benefit of combining edge computing and deep learning, where the best trade-off of the EdgeLoc system achieves 98.5\% localization accuracy within an average positioning time of only 2.31 ms.
\end{itemize}

The remainder of this paper is organized as follows.
In Section~\ref{related_work},
we briefly review the most recent studies of edge computing, indoor localization, and capsule networks.
Then, we present the design of the CapsNet model for robust indoor localization in Section~\ref{capsule_net}.
Next, we introduce the architecture and system dataflow of the EdgeLoc system in Section~\ref{framework} and Section~\ref{system_dataflow}, respectively.
In Section~\ref{experiment},
we provide experimental studies with system implementations and further show extensive evaluations with comprehensive analysis.
Finally, we conclude this work in Section~\ref{conclusion}.

\section{Related Work}\label{related_work}
In this section, we review the latest literature in edge computing, indoor localization and capsule networks, respectively.

\subsection{Edge Computing for IoT applications}
Edge computing has been rising in recent years with the proliferation of the Internet of Things and the ubiquitous coverage of wireless networks.
Edge computing enables unprecedented capacities for performing computation-intensive and latency-critical tasks, including real-time indoor localization.
In a typical edge computing system, edge servers provide computation,
service caching, and storage capacity to mobile users within the radio access network (RAN).
For mobile users of resource-limited mobile devices, they can benefit from edge computing by accessing services with low latency and high bandwidth, thereby saving energy on their own devices and shortening the delay of rather requesting services from the remote central cloud server.
With the huge demands for indoor location-based services, edge-IoT assisted indoor localization frameworks become critical and necessary.
In this work, we put edge computing into practice by implementing a real-world edge-IoT framework, which combines wireless access points, mobile devices and the edge server for robust indoor localization.

\subsection{Indoor Localization with RSS Fingerprinting}
With the ubiquitous accessibility of WiFi APs in indoor space,
RSS fingerprinting has become one of the most promising methodologies for indoor localization.
Meanwhile, there are still some key issues~\cite{RN35} that need to be formally addressed when performing localization with RSS fingerprinting, such as multipath effects, signal fluctuations and localization accuracy.
To achieve a higher localization accuracy,
various machine learning-based methods have been developed for indoor localization.
These methods range from the shallow learning methods such as KNN~\cite{RN190}, SVM~\cite{RN186}, to DL methods such as DNN~\cite{felix2016fingerprinting}, SAE~\cite{RN61}, CNN~\cite{RN187}, RNN~\cite{bae2019large} and their combinations.
For instance,
Li~\etal~\cite{RN39} proposed a feature-scaling-based KNN algorithm to assign differential weights to signal differences at different RSS levels and improved the localization accuracy.
In an early attempt in deep learning for indoor localization,
Felix~\etal~\cite{felix2016fingerprinting} used DNN and DBN to increase the accuracy of location estimation and reduce generalization errors in the dynamic indoor environment.
Moreover, Zhang~\etal~\cite{RN61} tamed the variant and unpredictable RSS signals for positioning with a four-layer DNN,
which was pre-trained by Stacked Denoising Autoencoder (SDA) to learn reliable features from noisy samples without hand-engineering.
Song~\etal~\cite{RN187} further proposed a scalable neural network model by combining CNN with SAE to deliver more accurate multi-building and multi-floor localization in different indoor environments and datasets.
More recently, RNN and Long Short-Term Memory have also been adopted to perform indoor localization with a sequential dataset to enhance the localization accuracy in large-scale indoor spaces~\cite{bae2019large}.
In this work, we use capsule networks for indoor localization to efficiently learn and preserve hierarchical representations from RSS fingerprints and thereby improving indoor localization accuracy.

\subsection{Capsule Networks}
The concept of `capsules' was firstly introduced by Hinton~\etal in~\cite{RN132},
where they used `capsules' to preserve the correlated spatial information of input data.
In a milestone work later,
Sabour~\etal~\cite{RN104} proposed CapsNet with dynamic routing for Capsules,
where the activity vector of each capsule is made up of several preset parameters,
such as position, orientation, scaling, and skewness.
With routing-by-agreement, the outcome of Capsules in higher levels can be predicted by the Capsules in lower levels.
Since then, many innovative works have been proposed based on CapsNet for different applications, including feature representation~\cite{RN119}, image classification~\cite{wang2019caps}, audio processing~\cite{RN128}
and multi-task learning~\cite{xiao2018mcapsnet}.
In~\cite{own2019signal}, Own~\etal used the SVM model to distinguish indoor environments and further employed conventional capsule networks for indoor localization with 2.4G and 5G WiFi signals, respectively.
In this work, we propose to employ CapsNet at the edge server for robust indoor localization with WiFi fingerprinting,
and the experimental results validate the effectiveness of CapsNet in extracting high-level features from WiFi fingerprinting.

\section{CapsNet Model for Indoor Localization}~\label{capsule_net}
\begin{figure}[t]
	\centering
	\includegraphics[width=0.47\textwidth]{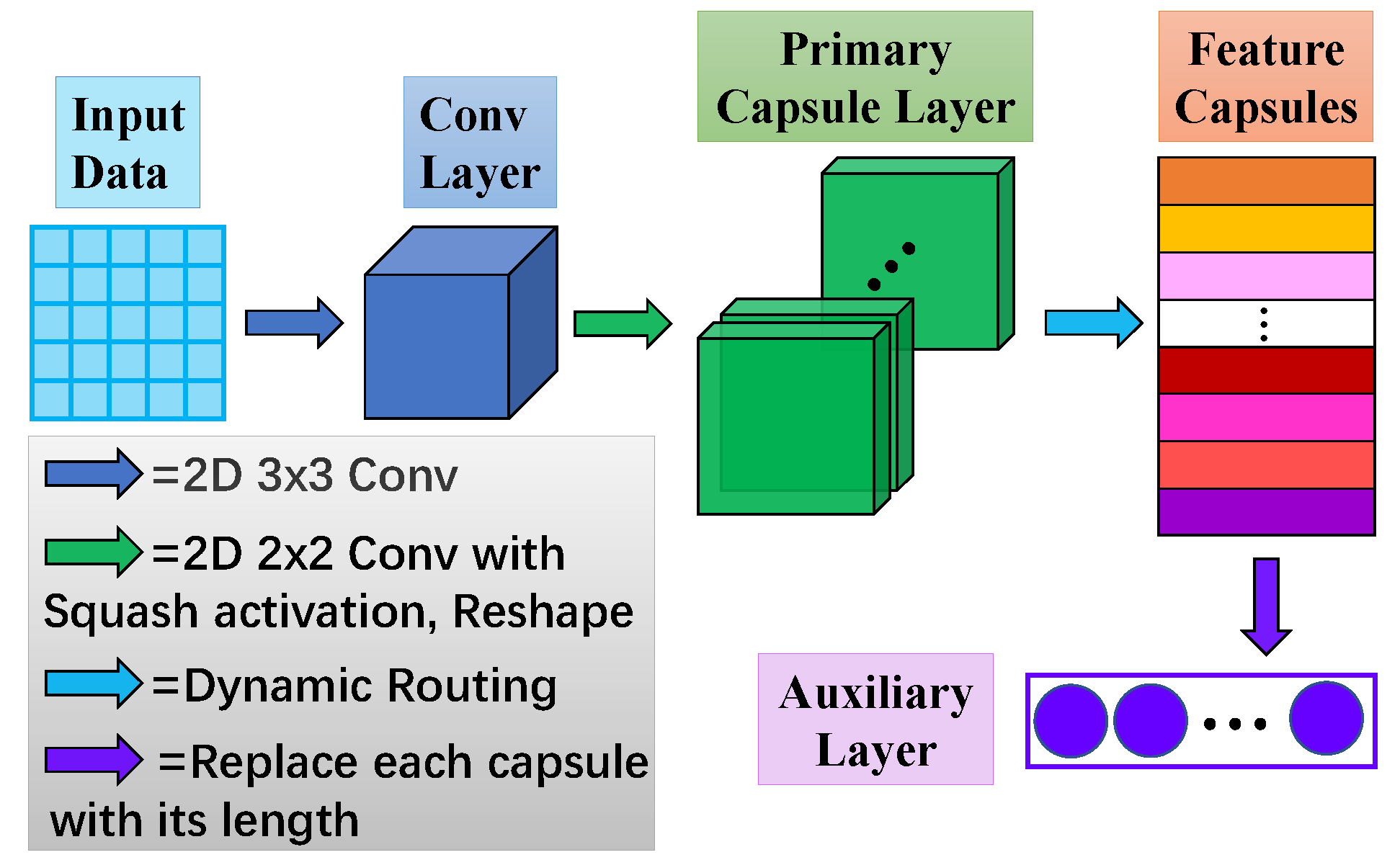}
	\caption{The architecture of CapsNet for indoor localization.}
	\label{CapsNet}
\end{figure}

As shown in Fig.~\ref{CapsNet}, the CapsNet model consists of five layers,
including the input data layer, a Convolutional (Conv) layer, a Primary-Capsule (PC) layer, a Feature-Capsule (FC) layer and the Auxiliary Layer.
The input fingerprinting data is processed by the convolutional operation with different filters in the Convolutional (Conv) Layer.
In the next Primary-Capsule (PC) Layer, the data is further processed by a Conv with squash activation and then reshaped.
After that, the processing of reshaped data is based on `Dynamic Routing' to derive the feature capsules in Feature-Capsule (FC) Layer.
At last, an Auxiliary Layer replaces each capsule of the FC Layer with its length to match the true label in the form of a One-Hot encoder~\cite{buckman2018thermometer}.
We introduce the core operations in the CapsNet as follows.

\subsection{Convolution Operation}
Let $x_i\in R$ be the one-dimensional data. The input data vector $\bm{x}^j$ of length $n$ is represented as $\bm{x}^j=[x^j_1,x^j_2,\dots,x^j_n],j=1,2,\dots M$, where $M$ is the number of training points and $n$ is the number of APs. A convolution operation involves a filter $\bm{w}^j\in \mathbb{R}^n$, which is applied to the vector $\bm{x}^j$ to produce a new feature. For instance, a feature $c^j_i$ is generated from the vector $\bm{x}^j$ by:
\begin{equation}
c^j_i=\mathnormal{f}(\bm{w}^j\centerdot \bm{x}^j+b^j).
\end{equation}
Here, $b^j\in \mathbb{R}^n$ is a bias term; $\mathnormal{f}$ is a non-linear activation function that introduces nonlinearities to CNN and is desirable for multi-layer networks to detect nonlinear features of input data.
The filter $\bm{w}^j$ is applied to every vector $\bm{x}^j$, where $j$ is the number of the vector to produce a feature map as:
\begin{equation}
\bm{c}^j=[c^j_1,c^j_2,\dots c^j_n].
\end{equation}

\subsection{Dynamic Routing}
A capsule is defined as a group of neurons in the CapsNet.
It is a vector that has both direction and length.
The direction of the capsule captures the entity's attributes.
The length of the capsule represents the probability of an entity's existence.
The shortcomings of CNNs are mostly related to the pooling layers.
As in CapsNet, these layers are replaced with a more appropriate criterion called `routing by agreement'.
Based on this criterion, the outputs are sent to all parent capsules in the next layer,
while their coupling coefficients are not equal.
Each capsule tries to predict the output of its parent capsules, and if this prediction corresponds to the actual output of the parent capsule, the coupling coefficient between these two capsules will increase.
The pseudocode of the dynamic routing of CapsNet is shown in Algorithm~\ref{routing}.

\begin{algorithm}[tb]
	\caption{Dynamic Routing Algorithm of CapsNet}
	\label{routing}
	\begin{algorithmic}[1]
		\Require
		The initialized output $\bm{u}_i$ of capsule $i$ in the Primary-Capsule layer, the number of iteration times $t$.
		\Ensure The output $\bm{v}_j$ of capsule $j$ in the Feature-Capsule layer.
		\State Initial the log-probability coupling coefficients $b_{ij}=0$.
		\For{the $t$-th iteration}
		 \For{all capsule $i$ in the PC layer}
		 \State $c_{ij}=\frac{exp(b_{ij})}{\sum_{j}exp(b_{ij})}$;
		 \EndFor
		 \For{all capsule $j$ in the FC layer}
		 \State $\widehat{\bm{u}}_{j|i}=\bm{W}_{ij}\bm{u}_i$;
		 \State $\bm{S}_j=\sum_{i}c_{ij}\widehat{\bm{u}}_{j|i}$;
		 \State $\bm{v}_j=\frac{||\bm{S}_j||^2}{1+||\bm{S}_j||^2}\frac{\bm{S}_j}{||\bm{S}_j||}$;
		 \EndFor
		 \For{all capsule $i$ in the PC layer and capsule $j$ in the FC layer}
         \State $a_{ij}=\boldsymbol{v}_{j}\hat{\boldsymbol{u}}_{j\mid i}$;
		 \State $b_{ij}=a_{ij}+b_{ij}$
		 \EndFor
		\EndFor \\
		\Return $\bm{v}_j$.
	\end{algorithmic}
\end{algorithm}

Consider $\bm{u}_i$ as the output of capsule $i$, its prediction for the parent capsule $j$ is computed as:
\begin{equation}
\widehat{\bm{u}}_{j|i}=\bm{W}_{ij}\bm{u}_i,
\end{equation}
where $\widehat{\bm{u}}_{j|i}$ is the prediction vector of the output of the $j_{th}$ capsule at a higher level, computed by capsule $i$ of the PC layer, and $\bm{W}_{ij}$ is the weighting matrix that the CapsNet needs to learn in backpropagation. Based on the degree of conformation between the capsules in the layer below and the parent capsules, the coupling coefficients $c_{ij}$ can be calculated by using the following Softmax function:
\begin{equation}
c_{ij}=\frac{exp(b_{ij})}{\sum_{j}exp(b_{ij})},
\end{equation}
where $b_{ij}$ is the log probability of whether capsule $i$ should be coupled with capsule $j$, and it is set to $0$ at the beginning of the routing by agreement process.
Accordingly, the input vector to the parent capsule $j$ is calculated as:
\begin{equation}\label{parent}
\bm{S}_j=\sum_{i}c_{ij}\widehat{\bm{u}}_{j|i}.
\end{equation}

Finally, the following non-linear squashing function is used to prevent the output vectors of Capsules from exceeding and further form the final output of each Capsule according to the value of its initial vector as defined in Equation~\ref{parent}:
\begin{equation}
\bm{v}_j=\frac{||\bm{S}_j||^2}{1+||\bm{S}_j||^2}\frac{\bm{S}_j}{||\bm{S}_j||},
\end{equation}
where $\bm{S}_j$ is the input vector to Capsule $j$ and $\bm{v}_j$ is the output.
The log probabilities should be updated in the routing process according to the agreement between $\bm{v}_j$ and $\widehat{\bm{u}}_{j|i}$, based on the fact that if the two vectors agree, they will have a large inner product. Therefore, the agreement $a_{ij}$ is used for updating log probabilities $b_{ij}$ and coupling coefficients $c_{ij}$ is calculated as:
\begin{equation}
a_{ij}=\bm{v}_j\widehat{\bm{u}}_{j|i}.
\end{equation}

Each capsule $k$ in the FC layer is associated with a loss function $l_k$, which puts high loss value on capsules with long output instantiation parameters when the entity does not actually exist. The loss function $l_k$ is computed as:
\begin{equation}
l_k=T_kmax(0,m^+-||\bm{v}_k||)^2+\lambda(1-T_k)max(0,||\bm{v}_k||-m^-)^2,
\end{equation}
where $T_k$ is $1$ whenever class $k$ is actually present and is $0$ otherwise. Terms $m^+$, $m^-$, and $\lambda$ are the hyperparameters to be learned in the training process.

\begin{figure}[t]
		\centering
	   \includegraphics[width=0.48\textwidth]{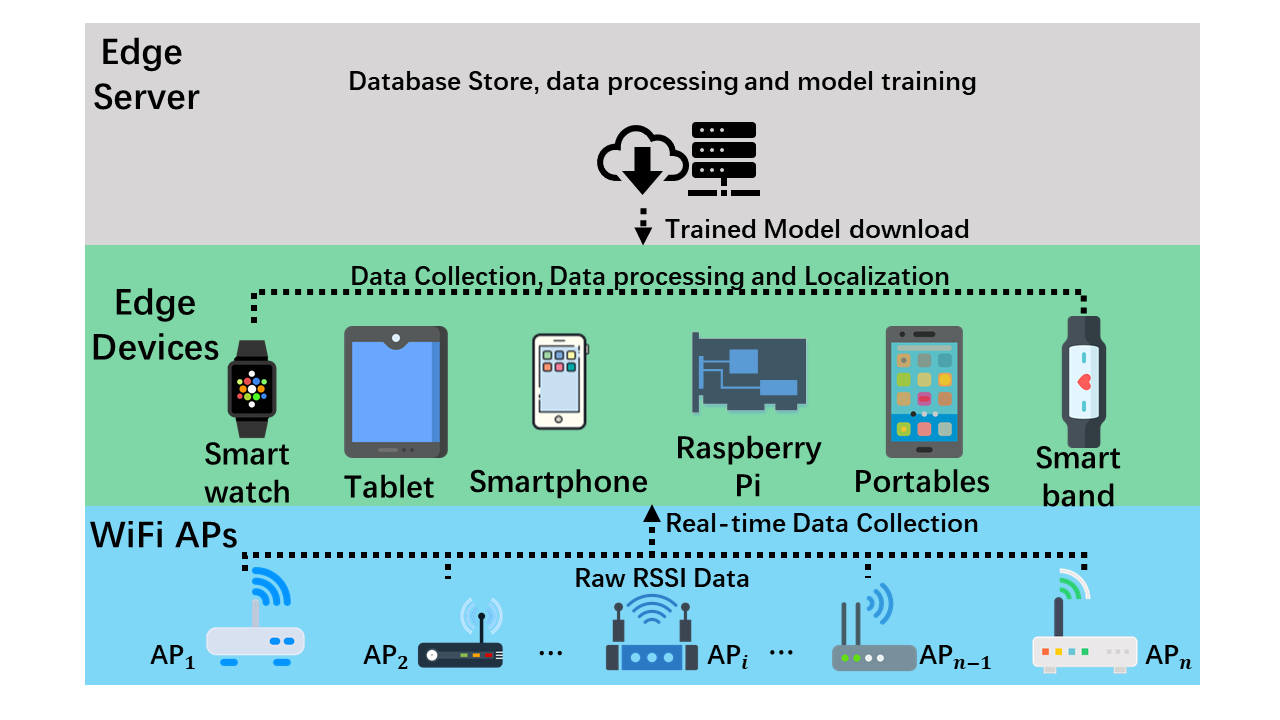}
	   \caption{The edge-IoT based framework for real-time indoor localization.}
	   \label{system}
\end{figure}

\section{The Architecture of Edge-IoT Framework}~\label{framework}
In this section, we first present an overview of the edge-IoT framework for robust indoor localization.

As shown in Fig.~\ref{system}, the architecture of EdgeLoc consists of three major components, a number of deployed WiFi APs, various edge devices of mobile users, and an edge server acting as the data processing unit. We introduce the details of each component as follows.
\begin{figure*}[t]
	\centering
	\includegraphics[width=0.6\textwidth]{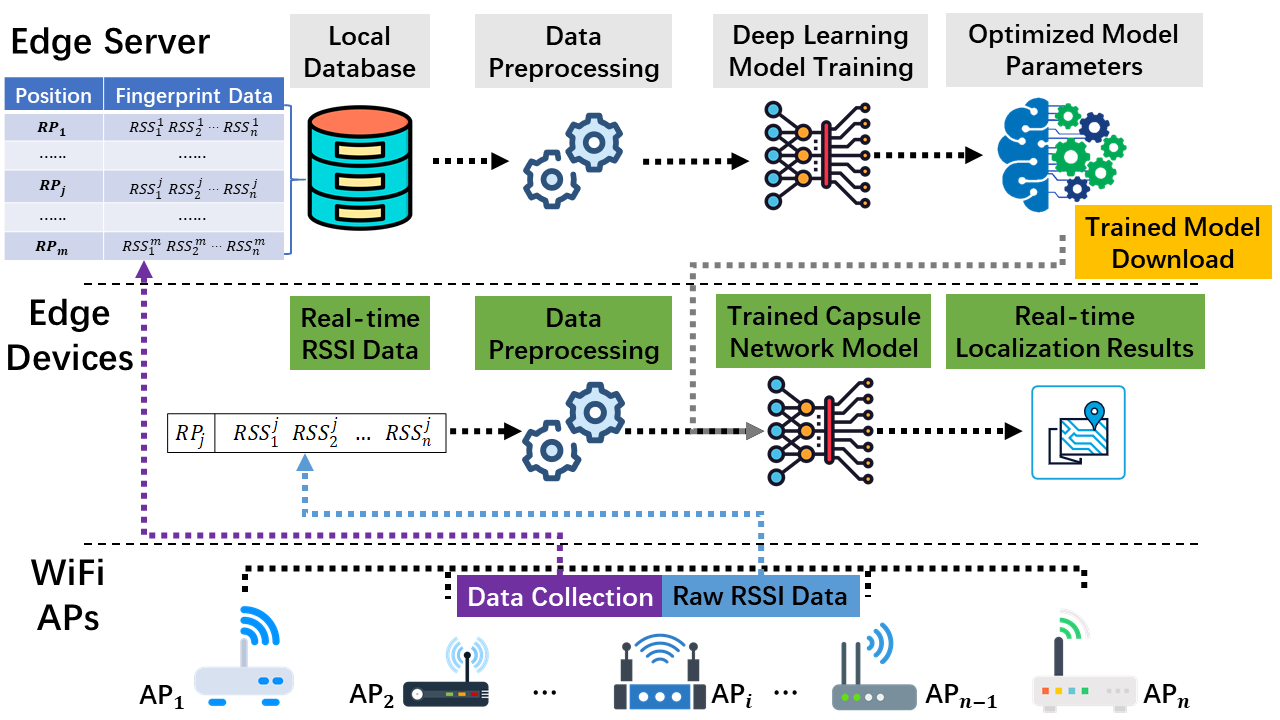}
	\caption{The dataflow of EdgeLoc based on the edge-IoT platform.}
	\label{EdgeLoc-Dataflow}
\end{figure*}

$1)$
\textbf{Edge Server:} The edge server contains a local database and a control center for the edge-IoT indoor localization system, by leveraging edge cloud computing to achieve data processing and model training.
Particularly, since the training processing of the CapsNet models is too complex for edge devices, the parameters of CapsNet are firstly learned through model training on the edge server, and then kept in the edge server.
When the local database is renewed, the parameter set will be fed into the training model for optimization.
For edge devices requesting localization services, the edge server will offload the trained deep learning model with the optimized parameters to them.
Thereafter, the edge IoT devices can directly run localization code and get the indoor localization results.

$2)$
\textbf{Edge Devices:} The portable IoT devices nowadays are increasingly ubiquitous and they are fully connectable to wireless networks, including WiFi.
A variety of wearable and portable devices can be leveraged for indoor localization, including the smartphones, smartwatches, tablets, and smart bands.
These edge devices generally have limited resources in computation and storage,
and they can only support primary computations, including data collection, data processing and localization using the trained CapsNet model.
Therefore, it is not feasible to perform the training process of models at the back-end of edge devices~\cite{RN165}.
As shown in Fig.~\ref{system}, in this work, we use a Raspberry Pi to represent resource-constrained IoT devices and conduct indoor localization experiments on it.
Correspondingly, we collect real-time data from WiFi APs and download the trained model from the edge server.

$3)$\textbf{Wireless Access Points:} The wireless APs are the fundamental components in the Edge-IoT framework for indoor localization. Generally, wireless APs are deployed at determined locations and all APs broadcast beacon frames to advertise their presence in the network (typically 100 ms per transmission).
Upon scanning the channels to receive beacon information from surrounding WiFi APs,
mobile devices further measure the RSS information of each AP by their equipped wireless cards~\cite{li2018multimodel}.
The RSS fingerprinting leverages the RSS values (\eg, a vector contains a series of RSS data) from multiple WiFi APs as a unique fingerprint of the determined location (\eg, reference point). With RSS fingerprints from different locations stored in a local database, the location of a user can be estimated by finding the best match of his RSS measurement and the fingerprints of the reference points~\cite{kim2018scalable}.

\section{The System Dataflow}~\label{system_dataflow}
In this section, we present the dataflow for real-time indoor localization processing based on the EdgeLoc system, as shown in Fig.~\ref{EdgeLoc-Dataflow}.
The data processing includes two computation phases: computations on edge devices and the edge server, respectively.

For computation at the edge server,
the RSS fingerprinting dataset is stored in the local database and then fed into the data preprocessing module.
After that, the preprocessed data is utilized for training the CapsNet model to derive the optimized model parameters.
For computation at edge devices,
the RSS fingerprinting data is first collected by edge devices from the surrounding WiFi APs and then sent to the data preprocessing module.
After that, the edge devices download the optimized model parameters from the edge server and use the trained CapsNet model for real-time localization.

\begin{figure*}[t]
	\begin{minipage}[t]{0.49\linewidth}
		\centering
		\includegraphics[width=0.98\textwidth]{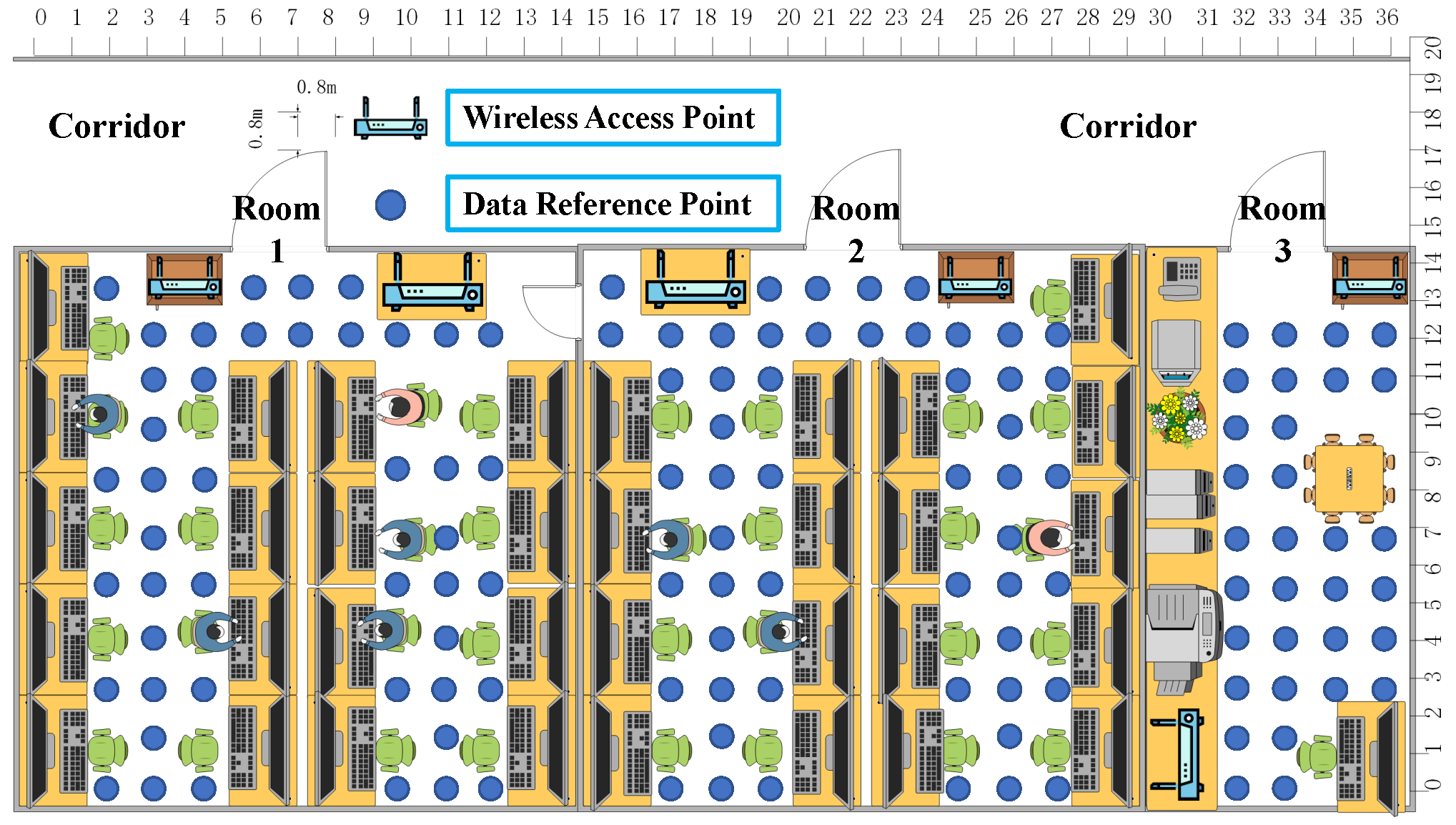}
		\caption{The floor plan of the experiment field with EdgeLoc.}
		\label{floor}
	\end{minipage}
	\hfill
	\begin{minipage}[t]{0.4\linewidth}
		\centering
		\includegraphics[width=0.98\textwidth]{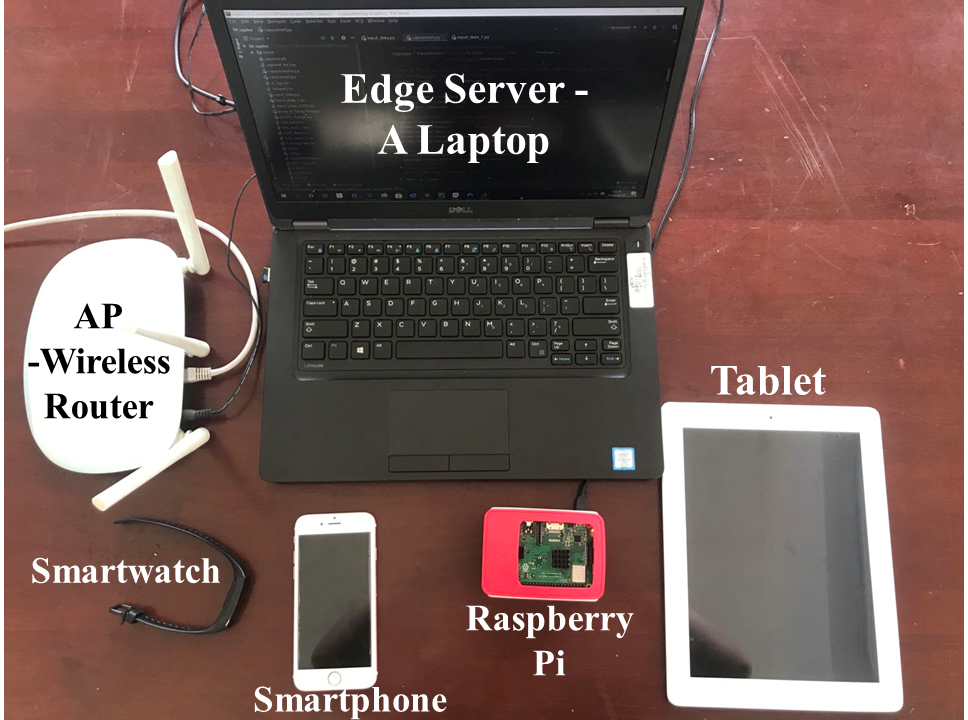}
		\caption{The illustration of system implementation in EdgeLoc.}
		\label{implementation}
	\end{minipage}
\end{figure*}

\subsection{RSS Fingerprinting Data}
We utilize RSS data collected from WiFi APs in an indoor environment to create an RSS fingerprint database.
Assuming that there are $n$ APs and $k$ reference points across the indoor entire space.
At each reference point, we collect $r$ pieces of RSS sampling data, which are labeled by the two-dimension location information (row and column) as the ground-truth.
In this way, the unrolled fingerprint database can be described as a huge matrix including $m\huge\times n$ vectors where $m=r*k$.

\subsection{Data Preprocessing}\label{preprocessing}
$a)$ RSS Data Processing:
To enrich the representations of RSS data,
we increase the dimension of original RSS fingerprints by adding a new feature set beyond itself. The detailed features sets are set as follow~\cite{RN108}:

$\bullet Raw$: The original features that are directly generated from the RSS readings.

$\bullet r$: The normalized the raw RSS readings, which is calculated by:
\begin{equation}
r_i=\begin{cases}
0&\text{$RSS_i$ is none},\\
0.1*(RSS_i-min)&
\text{otherwise},
\end{cases}
\end{equation}
where $r_i$ is the normalized RSS values from AP $i$, $RSS_i$ is the raw RSS values from AP $i$, $min$ is the lowest RSS value considering all of the fingerprints.

$\bullet R$: A set of features that represent the mutual differences of RSS data from different APs.
For instance, a basic entry of $r_i-r_j$ ($i,j\in n$) can represent the difference between the RSS values from AP $i$ and AP $j$.
Consequently, the overall feature matrix $R$ can be formulated as:
\begin{equation}
\bm{R}=
\left[
\begin{array}{ccccc}
0&r_1-r_2& r_1-r_3 &\cdots & r_1-r_n\\
r_2-r_1&0& r_2-r_3 &\cdots& r_2-r_n\\
r_3-r_1&r_3-r_2& 0 &\cdots& r_3-r_n\\
\vdots & \vdots  & \vdots & \ddots & \vdots \\
r_n-r_1&r_n-r_2& r_n-r_3 &\cdots  &0
\end{array}
\right].
\end{equation}

$b)$ Label Preprocessing:~\label{vector}
To determine the label of RSS fingerprints at each reference point,
we first divide the localization area into a number of small grids,
where each zone is a square area of $1.6\huge\times 1.6$ ${m^2}$.
To generate the label for each grid,
we adopt the One-Hot Encoding~\cite{buckman2018thermometer} to map each grid into a one-hot vector.
Consequently, each individual grid represents a categorical variable and the indoor localization task essentially becomes a classification problem across all grids with the fingerprints.

\subsection{Dataflow in model training}
We illustrate the layers of CapsNet in Fig.~\ref{dataflow} and introduce the dataflow over the CapsNet model from input to the output as follows.

$\bullet$ The input layer takes the feature matrix $R$ into the model, which is down-sampled to $n\huge\times n$.

$\bullet$ The second layer is a convolutional layer, where the size of the convolution kernel is $3\huge\times 3$ and the stride is $1$. In addition, the number of filters in this layer is to be learned.

$\bullet$ The next layer is the Primary Capsule layer. Similarly, the kernel for the convolutional operation of this layer is $3\huge\times 3$ and the stride is $2$. The number of channels (\ie, filters) and the dimension of the capsule in this layer are to be learned.

$\bullet$ The followed layer is the Feature (Digit) Capsules layer, where the number of capsules is the number of the grids in the localization area (illustrated in Section~\ref{vector}). The dimension of the capsule in this layer is the same as the one in the Primary Capsule layer, which is to be learned.

$\bullet$ The final layer is the output layer, which replaces each capsule with its length. The dimension of the output is the same as the one in the Feature Capsule layer.

\section{Experimental Studies}~\label{experiment}
\subsection{Experimental Setup}
We implement a real-world indoor localization system of EdgeLoc in an IoT lab in the main building of Beijing University of Posts and Telecommunications (BUPT) as illustrated in Figure~\ref{floor}.
To achieve the most efficient localization,
we deploy $6$ WiFi APs in an indoor area of 460${m^2}$,
covering three lab rooms (each room has two APs) along with a corridor area.
Here, all APs are TP-Link wireless routers.
Meanwhile, we set a number of reference points that evenly distributed across the floor space of each room, with the distance between two adjacent RPs as $0.8 m$.

\textbf{System Setup.}
As illustrated in Fig.~\ref{implementation},
we implement the EdgeLoc system by using a DELL Latitude 5480 laptop as the edge server and a Raspberry Pi 3 as the edge device.
The edge server laptop is equipped with a 4-thread Intel i7-7600U CPU of 2.9 GHz and 16GB RAM, and the edge device is equipped with a 64-bit quad-core ARMv8 CPU.
The localization model of EdgeLoc is implemented on Keras of TensorFlow using Python 3.6.

\textbf{Data Collection.}
In a real-world indoor space, the multi-path effects and fluctuations of RSS signals would always influence the accuracy and stability of indoor localization. In the EdgeLoc system, we employ a laptop installed with Phoenix Wi-Fi Collector to collect and store raw WiFi fingerprints.
To tame the variations in RSS signals,
we sample the RSS fingerprints of $6$ APs at each RP for $300$ times.
These samples are further stored in a local database for training and testing purposes.
Overall, we have collected $33,600$ data points,
with each data point labeled by its location in the form of row and column,
as shown in Fig.~\ref{floor}.
Thereafter, we split the dataset into two parts, with $80\%$ data points for training and the other $20\%$ testing.

\textbf{Baseline Methods.}
To comprehensively evaluate the performance of EdgeLoc,
we adopt four representative baseline methods of indoor localization for comparison as follows.
FS-$k$NN~\cite{RN39}: A feature-scaling-based $k$-nearest neighbor algorithm that assigns different weights to the signal differences at different RSS levels for localization.
SVM~\cite{RN186}: A multi-class support vector machines (SVMs) based indoor localization method.
CNN~\cite{mittal2018adapting}: A convolutional neural network-based indoor localization framework with RSS fingerprints.
CNNLoc~\cite{RN187}: A novel indoor localization framework using combine stacked auto-encoder (SAE) and convolutional neural network (CNN) for multi-building and multi-floor indoor localization.

\begin{figure}[t]
	\centering
	\includegraphics[width=0.49\textwidth]{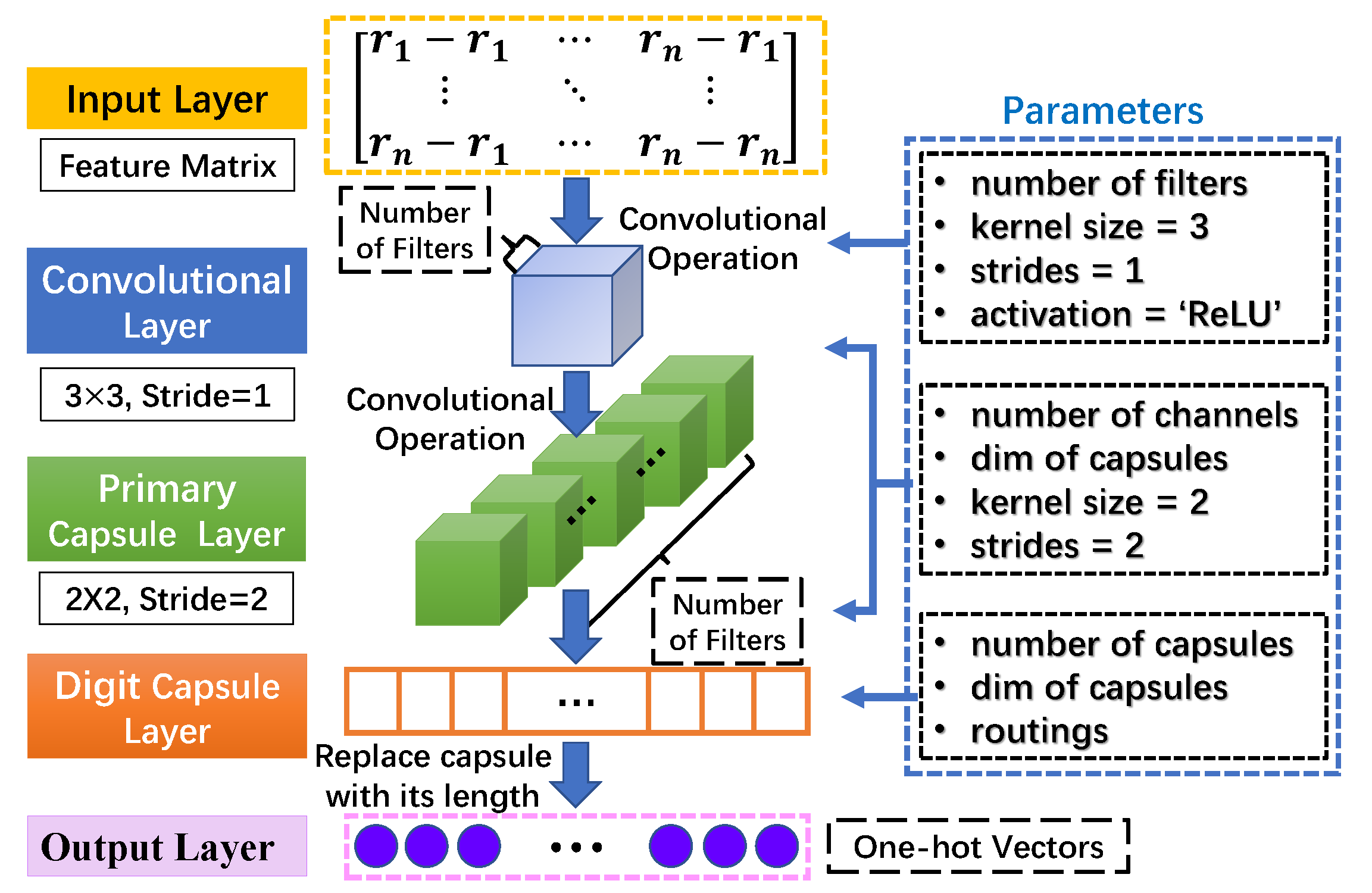}
	\caption{The dataflow and corresponding parameters of the CapsNet model.}
	\label{dataflow}
\end{figure}

\textbf{Parameter Settings.}
The basic dataflow of EdgeLoc for indoor localization is presented in Figure~\ref{dataflow},
where the EdgeLoc in EdgeLoc consists of five layers:
an input layer, a convolution layer, a primary capsule layer, a digit capsule layer and an output layer.
The input and output data of each layer are all specified in tensor format,
and the input and any layer share the tensor with the same shape.
The basic parameters of each individual layer are listed in the Parameters blocks.
For the Conv1 layer, we set the convolutional kernel size as 3, convolutional strides as 1, `ReLU' as the activation function, and further evaluate the impact of \emph{the number of filters (n\_filters)}.
For the Primarycap layer, we set the convolutional kernel size as 2, convolutional strides as 2, `Squash' as the activation
function, and further evaluate the impact of \emph{the number of channels (n\_channels)} and \emph{the dimension of the capsule (dim\_capsule)}.
For Digitcaps (the feature layer), we set the number of routing iterations as 3 and the number of capsules as the number of grids of the area, and further evaluate the impact of \emph{dim\_capsule}.
Note that we set the same number of \emph{dim\_capsule} for both Primarycap and Digitcaps.

\begin{figure}[t]
	\centering
	\subfloat [The CDF distribution of localization errors]
	{
		\label{CDF}
		\begin{minipage}[t]{0.85\linewidth}
			\centering
			\includegraphics[width=0.99\textwidth]{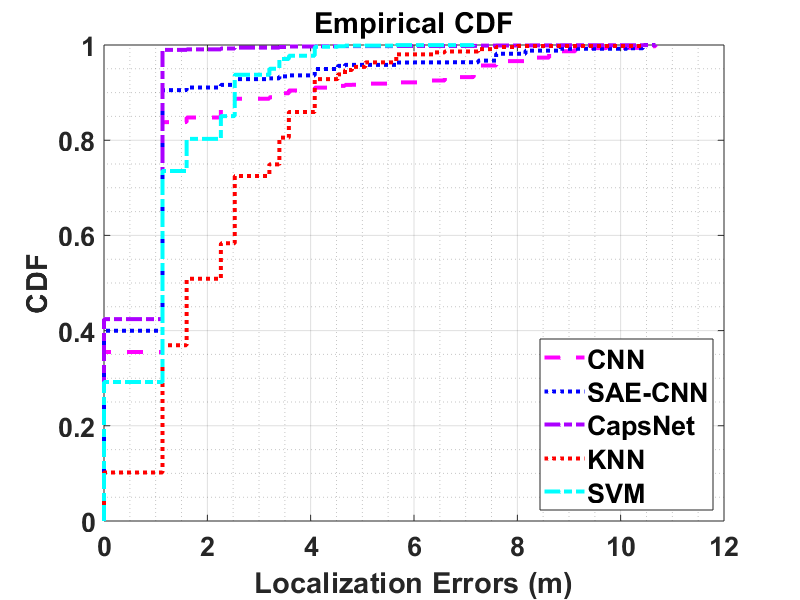}
			\par\vspace{0pt}
		\end{minipage}
	}
	\hfill
	\subfloat [The box-plot distribution of localization errors]
	{
		\label{box}
		\begin{minipage}[t]{0.85\linewidth}
			\centering
			\includegraphics[width=0.99\textwidth]{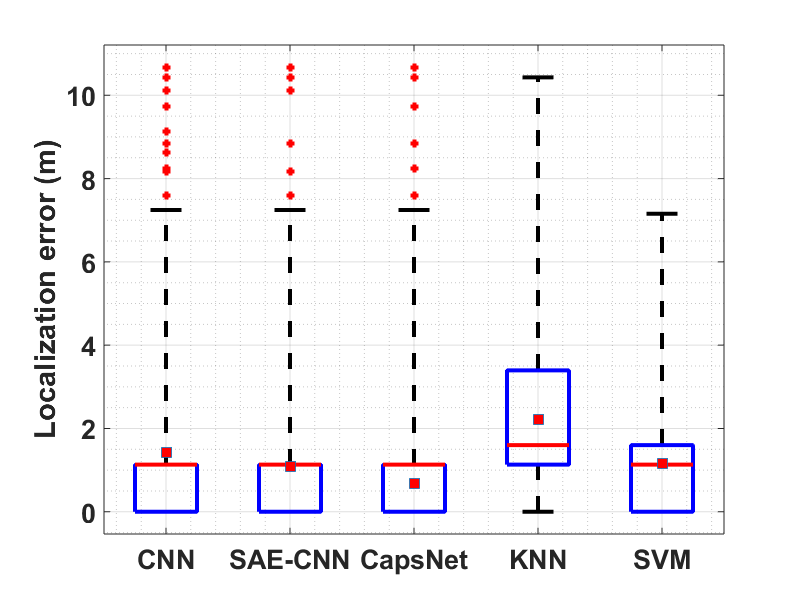}
			\par\vspace{0pt}
		\end{minipage}
	}
	\caption{The comparison results with baseline methods in CDF and boxplot.}
	\label{baseline_comparison}
\end{figure}

\begin{figure*}[t]
\centering
\begin{minipage}[t]{0.29\linewidth}
    \centering
    \includegraphics[width=0.99\textwidth]{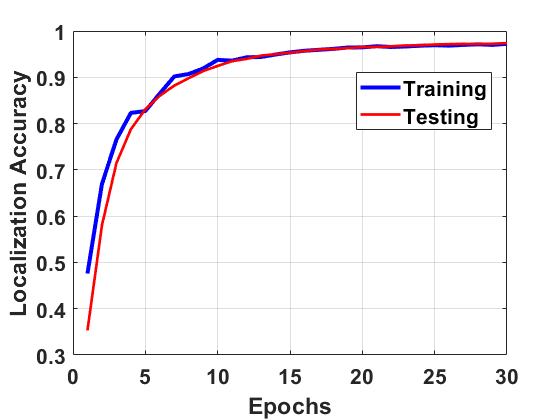}
    \par\vspace{0pt}
    \caption{Localization accuracy with training dataset and testing dataset.}
    \label{model-log}
\end{minipage}
    \hspace{0.01\linewidth}
\begin{minipage}[t]{0.6\linewidth}
	\subfloat [Primarycap layer with 8 channels]
	{
		\label{acc1}
		\begin{minipage}[t]{0.45\linewidth}
			\centering
			\includegraphics[width=0.99\textwidth]{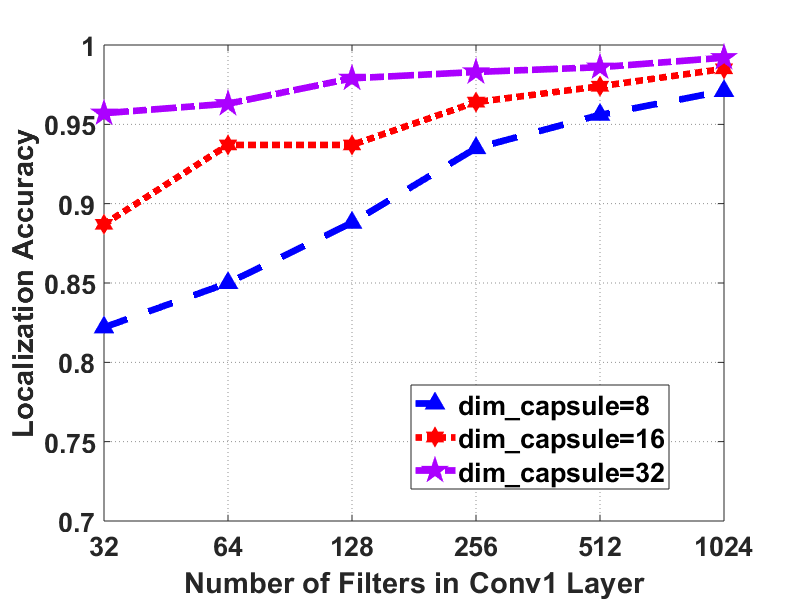}
			\par\vspace{0pt}
		\end{minipage}
	}
	\hfill
	\subfloat [Primarycap layer with 16 channels]
	{
		\label{acc2}
		\begin{minipage}[t]{0.45\linewidth}
			\centering
			\includegraphics[width=0.99\textwidth]{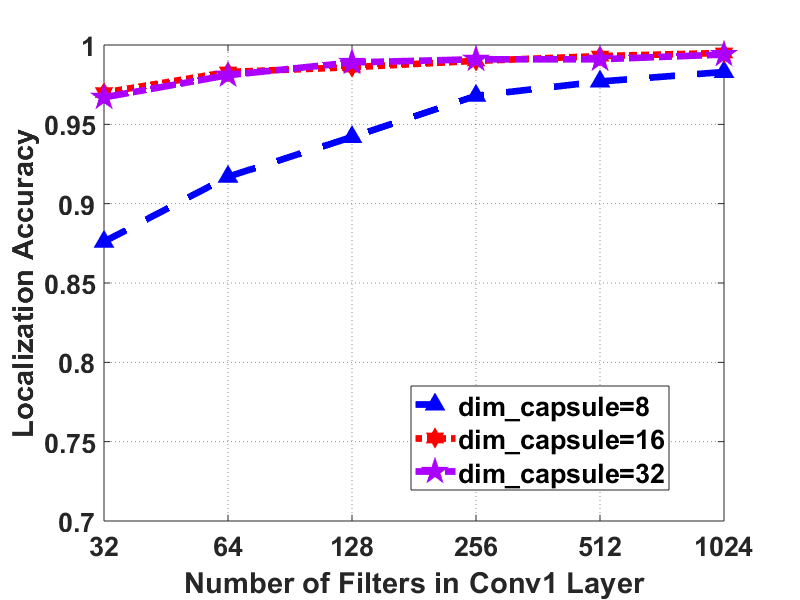}
			\par\vspace{0pt}
		\end{minipage}
	}
	\caption{Comparison of localization accuracy by EdgeLoc with 8 channels and 16 channels}
	\label{acc}
\end{minipage}
\end{figure*}

\begin{figure*}[t]
\centering
\begin{minipage}[t]{0.29\linewidth}
    \centering
    \includegraphics[width=0.99\textwidth]{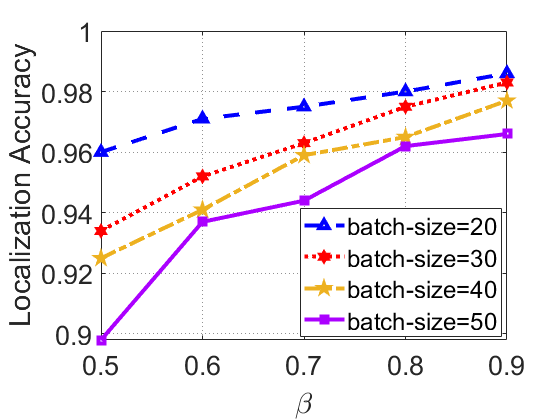}
    \par\vspace{0pt}
    \caption{Comparison of localization accuracy by EdgeLoc with different sizes of batch and training samples.}
    \label{rate}
\end{minipage}
    \hspace{0.01\linewidth}
\begin{minipage}[t]{0.6\linewidth}
	\centering
	\subfloat [Primarycap layer with 8 channels]
	{
		\label{time1}
		\begin{minipage}[t]{0.45\linewidth}
			\centering
			\includegraphics[width=0.99\textwidth]{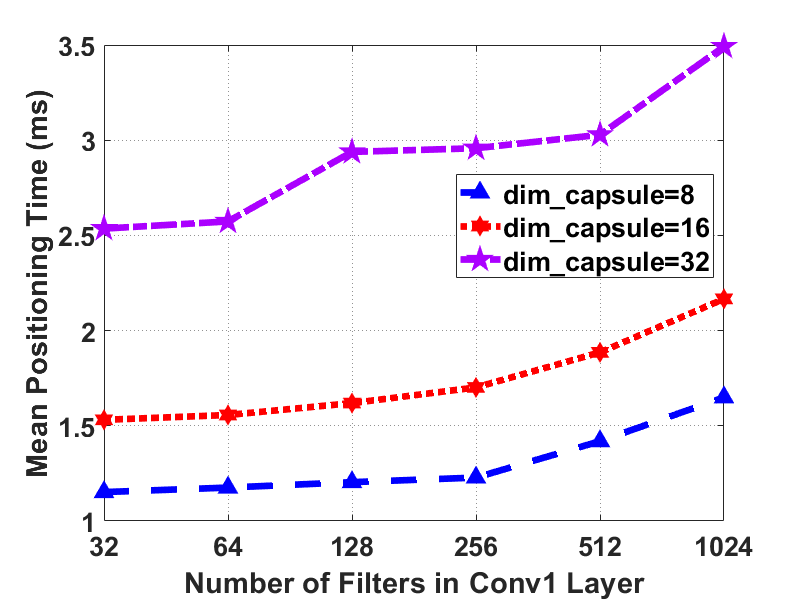}
			\par\vspace{0pt}
		\end{minipage}
	}
    \hfill
	\subfloat [Primarycap layer with 16 channels]
	{
		\label{time2}
		\begin{minipage}[t]{0.45\linewidth}
			\centering
			\includegraphics[width=0.99\textwidth]{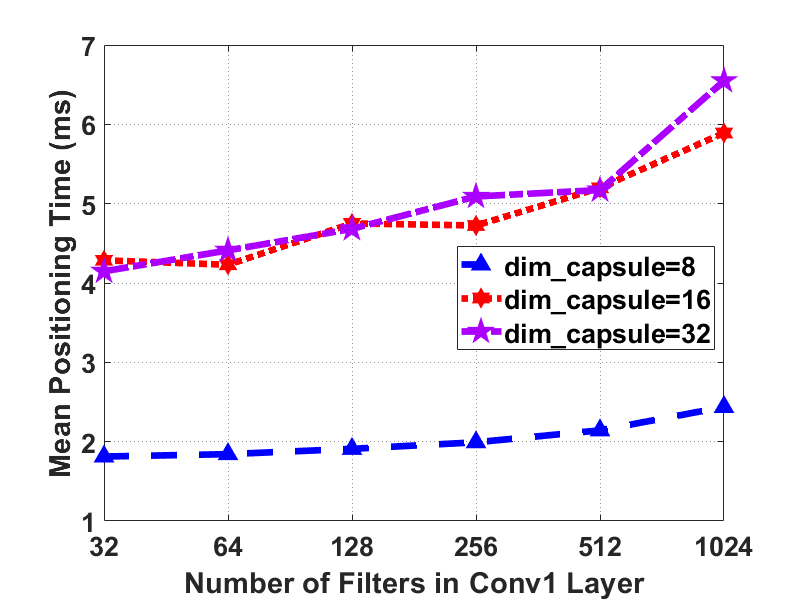}
			\par\vspace{0pt}
		\end{minipage}
	}
    \caption{Comparison of positioning time by EdgeLoc with 8 channels and 16 channels.}
    \label{time}
\end{minipage}
\end{figure*}

\subsection{Performance on Real-World Indoor Localization}
We present the performance of EdgeLoc and other baseline methods in Fig.~\ref{baseline_comparison}.
First, the distribution of localization errors by different methods is revealed in Fig.~\ref{CDF}, where the FS-$k$NN shows unsatisfactory localization performance with more than 60\% testing results containing errors from 2$m$ up to 8$m$.
For the SVM-based localization method, there are still 20\% testing results with errors larger than 2$m$ and 5\% testing results with errors larger than 3$m$.
Moreover, CNN-based localization approaches extract high-level features and thereby enhancing the localization accuracy, with 80\% testing sample having errors less than 2$m$).
As it can be observed, the SAE-CNN further improves the performance of CNN,
by encoding raw RSS fingerprints into features that act as the inputs of CNN.
Since EdgeLoc preserves the valuable spatial information for between-layer neurons,
it successfully achieves the best indoor localization results over all baseline methods,
with 99\% testing results are within errors lower than 2$m$ and over 40\% testing samples are with errors around 1$m$.

We further provide a box-plot for localization results in Fig.~\ref{box}.
Obviously, the KNN method performs the worst with the largest Interquartile Range (IQR, \ie, the distance between first quartile and third quartile) of localization errors.
In comparison with CNN and SAE-CNN, the proposed EdgeLoc has a smaller IQR and a much lower average localization error at 0.68$m$.
The above results have shown that EdgeLoc outperforms the conventional machine learning methods and the state-of-the-art deep learning methods in indoor localization with RSS fingerprints.

\textbf{Parameter's impact on localization accuracy:}
Before tuning parameters of the CapsNet model,
we visualize the process of model training in Fig.~\ref{model-log},
where the CapsNet model achieves over 90\% localization accuracy after 8 epochs for both training data and testing data.

In tuning the parameters, for the Primarycap layer, we set the number of channels (\ie, \emph{n\_channels}) as 8 and 16 for evaluation studies, respectively.
We tune the dimension of the output vectors by the capsules (\ie, \emph{dim\_capsule})
from 8 to 16, and 32.
Moreover, we conduct a grid search for the number of filters ($n\_filters$) in the Conv1 layer over the set of {32, 64, 128, 256, 512, 1024}.
The evaluation results of the parameter's impact on localization accuracy are presented in Figure~\ref{acc}.

First, the overall localization accuracy of all methods improves steadily with a larger \emph{n\_filter} in the Conv1 layer.
Meanwhile, as the \emph{n\_channels} increases from 8 to 16, the localization accuracy only shows slight enhancement.
Second, the \emph{dim\_capsule} has a direct impact on the indoor localization accuracy,
since that EdgeLoc shows an accuracy improvement of up to 10\% in both Fig.~\ref{acc1} and Fig.~\ref{acc2}.
Meanwhile, with the increasing \emph{n\_filters} in the Conv1 layer, the accuracy improvement generally shrinks to below 5\%.
The above evaluations give us an insight into the contributions of different components in EdgeLoc, where the best trade-off of EdgeLoc is with 64 filters in the Conv1 layer, 8 capsules and a dimension of 16 in each capsule.

Fig.~\ref{rate} shows the localization accuracy of EdgeLoc with different sizes of training samples as well as different sizes of the batch during the training processing.
Here, we use $\beta$ to denote the size of training samples in proportion to the overall RSS fingerprinting dataset.
As revealed by the experimental results, with the larger size of training samples in the whole dataset, the localization accuracy will be higher.
Moreover, take $\beta=0.5$ for an example, the performance of EdgeLoc decreases from 0.96 to 0.9 with the increasing batch sizes from 20 to 50.

\textbf{Parameter's impact on the mean positioning time:}
The average positioning time would directly influence the user experience of an indoor localization system.
Therefore, in this evaluation,
we explore the impact of parameters on the mean positioning time by the edge server of the proposed EdgeLoc system.
Similar to evaluations on localization accuracy,
we set the \emph{n\_channels} of the Primarycap layer as 8 and 16, respectively,
with other parameters using the same tuning spaces.

As shown in Fig.~\ref{time1} and Fig.~\ref{time2},
the positioning time generally increases with larger \emph{n\_channels} in the Primarycap layer, larger \emph{dim\_capsule} and larger \emph{n\_filters} in the Conv1 layer.
In addition, Table~\ref{time3} shows the average positioning time in ms v.s. batch size, where EdgeLoc is with 64 filters in the Conv1 layer, 8 capsules and the \emph{dim\_capsule} of 16 at the edge. When the batch size becomes larger, the mean positioning time is reduced from 2.31 ms to 1.82 ms.
Therefore, by jointly considering the performance of localization accuracy in Fig.~\ref{acc} and mean positioning time in Table~\ref{time3},
we find that the EdgeLoc with 1024 filters in the Conv1 layer, 8 capsules and 16 in \emph{dim\_capsule} can achieve the best trade-off between localization accuracy and positioning time at the edge server (\eg, 98.5\% accuracy within an average time of 2.31 ms).

\begin{table}[t]
	\centering
	\caption{The mean positioning time for different batch sizes.}
	\label{time3}
	\begin{tabular}{c|c|c|c|c}
		\toprule
		Batch Size &20&30&40&50 \\
		\midrule
		Mean Positioning Time (ms) &2.31&2.10&1.96&\textbf{1.82}\\
		\bottomrule
	\end{tabular}
\end{table}

\subsection{Extensive Experiments on UJIIndoorLoc Dataset}
To further validate the scalability of EdgeLoc,
we evaluate it on the UJIIndoorLoc dataset~\cite{torres2014ujiindoorloc}.
The UJIIndoorLoc dataset covers three different buildings (with ID 0, 1, and 2) of more than 110,000 ${{\rm{m}}^{\rm{2}}}$ indoor areas, with totally 19,937 training samples and 1111 test samples, respectively.
We choose Building 0 from UJIIndoorLoc as the target for localization and select top-40 APs (out of total 520 APs) to characterize RSS fingerprints, by ranking all APs' frequency of occurrence in descending order.
The localization performance of all methods is presented in Table~\ref{average1},
where EdgeLoc achieves the best localization performance across all different floor levels in Building 0.

\begin{table}[t]
	\centering
	\caption{The average localization errors (m) by different algorithms for Building 0 of UJIIndoorLoc Dataset.}
	\label{average1}
	\begin{tabular}{c|c|c|c|c|c}
		\toprule
		\diagbox[width=5em,trim=l]{Models}{Level}&0&1&2&3& all \\
		\midrule
		EdgeLoc &\textbf{8.28}&\textbf{7.36}&\textbf{8.32}&\textbf{7.70}&\textbf{7.90}\\
		\midrule
		KNN &8.43&7.50&8.61&7.86&8.10 \\
		\midrule
		SVM &8.85&7.68&9.49&8.14&8.54 \\
		\midrule
		CNN &8.59&8.15&8.85&7.79&8.35 \\
		\midrule
		SAE-CNN &8.43&8.27&9.12&9.05&8.72 \\	
		\bottomrule
	\end{tabular}
\end{table}

\textbf{Impact of the number of WIFI APs on the localization accuracy for UJIIndoorLoc dataset:}
Table~\ref{average2} shows the average localization error of EdgeLoc by using different numbers of APs in UJIIndoorLoc dataset for indoor localization.
We first rank the 520 APs in descending order with their frequency of occurrence and select the top 20, 30 ,40 and 50 APs to generate fingerprint data.
As revealed in Table~\ref{average2}, the larger the number of involved APs is,
the higher the accuracy of indoor localization will be. In addition, the performance of EdgeLoc converges with 40 APs for localization, which demonstrates that the RSS fingerprints from top-40 APs are already sufficient for localizing mobile users.

\begin{table}[ht]
\centering
\caption{The average localization error (m) of EdgeLoc based on different numbers of WiFi APs in Building 0.}
\label{average2}
  \begin{tabular}{c|c|c|c|c|c}
		\toprule
		\diagbox[width=5em,trim=l]{APs}{Level} &0&1&2&3& all \\
		\midrule
		50 APs &8.63&7.53&\textbf{8.26}&\textbf{7.70}&8.03 \\
		\midrule
		40 APs &\textbf{8.28}&\textbf{7.36}&8.32&\textbf{7.70}&\textbf{7.90} \\
		\midrule
		30 APs &9.52&8.39&8.58&9.32&8.95 \\
		\midrule
		20 APs &10.61&9.30&8.98&9.36&9.56 \\
		\bottomrule
  \end{tabular}
\end{table}

\section{Conclusion}~\label{conclusion}
In this work, we have proposed EdgeLoc,
an Edge-IoT framework for robust indoor Localization using capsule networks.
We develop CapsNet based on the state-of-the-art capsule networks, and to the best of our knowledge, we are the first to employ CapsNet for indoor localization.
A prototype system of EdgeLoc has been further set up in a real-world experiment field.
The extensive experimental studies have shown the success of bridging edge computing and deep learning for indoor localization, as EdgeLoc achieves up to 98.5\% accuracy for indoor localization at an average positioning time of only 2.31 ms.

\section{Acknowledgment}
The preliminary version of this article was published in the 2020 IEEE International Conference on Communications (ICC 2020).

\bibliographystyle{IEEEtran}
\bibliography{edgeloc}

\end{document}